\documentclass[
 reprint,
 preprintnumbers,
 amsmath,
 amssymb,
 aps,
 prstab,
 floatfix,
]{revtex4-1}

\usepackage{graphicx}			
\usepackage{verbatim}			
\usepackage{wasysym}			
\usepackage{mathtools}			
\usepackage[makeroom]{cancel}		
\usepackage{dcolumn}			
\usepackage{hyperref}			
\usepackage{color}			

\newcommand{\pd}{\partial}				

\newcommand{\omsc}{\omega^\mathrm{sc}}
\newcommand{\omsci}{\omega^\mathrm{sc}_i}

\newcommand{\omxi}{\omega_{x i}}
\newcommand{\domxi}{\delta \omega_{x i}}

\newcommand{\omx}{\omega_{x}}
\newcommand{\omc}{\omega^\mathrm{c}}
\newcommand{\oms}{\omega_\mathrm{s}}

\newcommand{\omi}{\omega_{i}}
\newcommand{\domi}{\delta \omega_{i}}
\newcommand{\domo}{\delta \omega_{0}}

\newcommand{\xbar}{\bar{x}}

\begin{document}

\title{Space Charge Effects for Transverse Collective Instabilities in Circular Machines}
\author{A.~Burov}
\email{burov@fnal.gov}
\affiliation{Fermilab, PO Box 500, Batavia, IL 60510-5011}
\date{\today}

\begin{abstract}

A brief historical review is presented of progressing understanding of transverse coherent instabilities of charged particles beams in circular machines when both Coulomb and wake fields are important. The paper relates to a talk given at ICFA Workshop on Mitigation of Coherent Beam Instabilities in Particle Accelerators, 23-27 September 2019 in Zermatt, Switzerland.

\end{abstract}


\maketitle

 



\section{\label{sec:Int}Introduction}

Fifty years ago, the first significant publication was presented on transverse collective instabilities of space-charge-dominated beams in circular machines; it was a CERN preprint of G.~Merle and D.~M\"{o}hl ``The stabilizing influence of nonlinear space charge on transverse coherent oscillations''~\cite{Merle:1969zz}. A relatively simple equation of motion was suggested there as something obvious. Although it was, strictly speaking, neither obvious nor even quite correct, as further studies have shown, it played and continues to play an extraordinarily important role. Thus, this anniversary adds a special flavor to the request of the workshop organizers to review the main results in this area of beam dynamics.  

Purporting to fulfill that, this paper is divided in two sections, on coasting and bunched beams respectfully. We rather rarely deal with coasting beams in circular machines, but still they deserve a special attention not only for themselves~\cite{Prost:2011zza, LebECooler}  but also as relatively simple configurations to start from and get some key ideas. This set of ideas includes a concept of rigid slices and strong space charge as its justification. Also, it includes interplay of Landau damping (LD), space charge (SC) and octupoles, showing the importance of their polarity, in particular. These ideas, common for coasting and bunched beams, are presented in Sec.~\ref{SecCoast} and used in Sec.~\ref{SecBunch}. In the latter section, SC-modification of the transverse mode coupling instability (TMCI) is discussed, including paradoxes which were resolved in a discovery of convective instabilities.        

The goal of this paper is to present, in a compact way, the main results in the area of beam dynamics, specified by the subject, where both SC and wake field are important. To a certain degree, such a task cannot be free from some subjectivity and arbitrariness, and I beg pardon of those colleagues who will find some valuable results underrepresented or not presented at all.

\section{\label{SecCoast} Coasting Beams}

To analyze the beam stability with SC, a linear equation of motion was suggested by G.~Merle and D.~M\"{o}hl in 1969~\cite{Merle:1969zz}:
\begin{equation}
\frac{d^2 x_i}{dt^2} + \omxi^2 x_i + 2 \omx \omsci (x_i -\bar{x}) +  2 \omx \omc \bar{x} =0 \,.
\label{MMEq}
\end{equation}
Here $x_i=x_i(t)$ is a transverse offset of a particle $i$, $\omxi$  is the betatron frequency of the particle $i$, $\omx$ is the average betatron frequency, $\xbar=\xbar(t)$ is an average offset of that beam {\it slice} where the particle $i$ is located at the given moment of time $t$, $\omsci <0$ is the SC frequency shift of the particle, and $\omc$ is the coherent frequency shift parameter, proportional to the ring impedance.  The full time derivative $d/dt$ is expressed through the partial ones, $d/dt = \pd/\pd t + \omi \pd/\pd \theta$, where $\omi$ is the revolution frequency of the particle, and $\theta =s/R$ is the azimuthal angle, with $s$ as the conventional longitudinal coordinate and $R$ as the average ring radius. The term `slice' refers to the group of beam particles which Coulomb fields affect the given particle number $i$, i.e. the particles with positions somewhere between $s_i -a/\gamma$ and $s_i +a/\gamma$, where $a$ is the beam transverse size and $\gamma$ is the Lorentz factor.   

Equation~(\ref{MMEq}) implies two important things. 

First, it implies that $x_i$ relates to the driven part of the single-particle oscillations, excited by the collective motion of the centroids $\xbar$, while constant amplitudes of free oscillations determine the space charge frequency shifts. That is why the offset $x_i$ is of the order of centroid offsets, $x_i \sim \xbar$, so it can be considered infinitesimally small, while the incoherent amplitudes are of the order of the beam transverse size.  

Second, this equation assumes that each beam slice oscillates as a rigid body, allowing for a representation of the SC force in the simple way it is done there. Because of this assumption, Merle-M\"{o}hl approach is sometimes addressed as the {\it rigid-slice} or {\it frozen-field} model. Possible incorrectness of this assumption, as well as its very existence, was realized much later, when some strange features of Eq.~(\ref{MMEq}) were discovered.  

For a coasting beam, eigenfunctions of Eq.~(\ref{MMEq}) have the form 
\begin{equation}
x_i\,, \xbar \propto \exp\left[-i\,(\omx + n\omega_0 + \omega)t + in \theta \right]\,
\label{eigen1}
\end{equation}
where $n$ is an arbitrary integer, $\omega_0$ is the average revolution frequency, and $\omega$ is a frequency shift of the mode $n$. Substitution of this form into Eq.~(\ref{MMEq}) yields for the complex amplitudes
\begin{equation}
x_i = \xbar \frac{\omc - \omsci}{\omega - \omsci -\domxi -n\domi}\,,
\label{eigen2}
\end{equation}
with the lattice frequency shifts $\domxi=\omxi -\omx$ and $\domi=\omi-\omega_0$. 
Note that without lattice frequency spread, $\domxi=\domi=0$, there is always the rigid-bunch solution, $x_i=\xbar$, with $\omega=\omc$, independently of the SC tune shifts $\omsci$. This important physical property of Eq.~(\ref{MMEq}) is a consequence of its SC representation by means of the term $\propto x_i - \xbar$. In fact, the Merle-M\"{o}hl equation is the only possible linear dynamic equation consistent with the given incoherent spectrum, its lattice and SC parts, which represents the coherent SC term by means of $\propto \xbar$ term only, preserving the rigid-bunch mode for zero lattice tune spread, as it must be from the first principles.  

By averaging over the particles, writing the sums as the phase space integrals with the distribution function, one gets the dispersion relation, i.e. the equation for the sought-after eigenfrequency $\omega$, 
\begin{equation}
1 = - \int d\Gamma \, J_x \frac{\pd f}{\pd J_x} \frac{\omc - \omsci}{\omega - \omsci -\domxi -n\domi +i o}
\label{DispEq}
\end{equation}
Here $f=f(J_x, J_y, \delta p/p)$ is the unperturbed distribution density as a function of transverse actions $J_{x,y}$ and the relative momentum offset $\delta p/p$, normalized to unity, $\int d\Gamma f =1$, where $d\Gamma = dJ_x\,dJ_y\, d\delta p/p$; the single-particle subscript $i$ has to be understood as indication to related functional dependences, i.e. $\domxi \rightarrow \delta \omega_x(J_x, J_y, \delta p/p)$, etc. To get Eq.~(\ref{DispEq}) from Eq.~(\ref{eigen2}), the Hereward rule~\cite{hereward1969landau} was used,
$$\sum_i (...) \rightarrow - \int d\Gamma \, J_x\, \pd f/\pd J_x \,(...) \,,$$  
and the Landau rule of going around the pole is explicitly marked, $\omega \rightarrow \omega + i o$, where $o$ is an infinitesimally small positive number. 

It is straightforward to see from the dispersion relation~(\ref{DispEq}) that without lattice frequency spreads, at $\domxi=\domi=0$, the eigenfrequency $\omega=\omc$, independently of the SC tune shift. Thus, even if the phase space density of the resonant particles were not zero, i.e. there were particles with the same tune as the coherent mode, still there would be no Landau damping (LD), irrespectively to nonlinearity of SC distribution.

The dispersion relation in the form~(\ref{DispEq}) was first derived by D.~M\"{o}hl and H.~Sch\"{o}nauer in 1974~\cite{Mohl:1974vj}, not in the original preprint of Merle and M\"{o}hl. In the latter, some mathematical mistakes were adopted, so the dispersion relation was derived incorrectly. Due to this, it was mistakenly concluded there that SC nonlinearities may contribute to Landau damping of coasting beams even without the lattice frequency spread. This mistake was later repeated in Ref.~\cite{reich1972transverse} and corrected by M\"{o}hl and Sch\"{o}nauer~\cite{Mohl:1974vj}. That is why it seems fair to call Eq.~(\ref{MMEq}) Merle-M\"{o}hl equation of motion and Eq.~(\ref{DispEq}) M\"{o}hl-Sch\"{o}nauer dispersion relation.


After the simplest case of no lattice frequency spreads, the next by simplicity is a two-stream beam, $\domi=\pm \domo$. For KV transverse distribution with a constant SC frequency shift, $\omsci=\omsc$, Eq.~(\ref{DispEq}) yields the following spectrum: 
\begin{equation}
\omega=\frac{\omc + \omsc}{2} \pm \sqrt{\frac{(\omc - \omsc)^2}{4} + n^2 \domo^2}
\label{2stream} 
\end{equation}
The instability is driven by the coherent tune shift $\omc$, so the most unstable modes are positive ones, i.e. those associated with the sign $+$ in Eq.~(\ref{2stream}). A more detailed analysis shows that for them the two streams of the beam oscillate approximately in phase, so their wakes add to each other. For strong SC, $|\omsc| \gg \max(|n|\domo,\, |\omc|)$, the spectrum of positive modes reduces to
\begin{equation}
\omega= \omc + n^2 \domo^2/\omsc \,, 
\label{2stream2} 
\end{equation}
being far away from the incoherent spectrum localized around $\omsc$. Thus, the negative modes not only are barely excited by the wake, but also stay close to the incoherent spectrum, so their stability would be provided automatically when the positive modes are stable. 

A very general method of analysis of integral dispersion relations like Eq.~(\ref{DispEq}) was presented in Ref.~\cite{Ruggiero:1968vf}. The idea was to reverse the problem: instead of finding the eigenfrequency $\omega$ when the coherent tune shift $\omc$ is given, let us do the opposite, find the coherent tune shift $\omc$ for a given eigenfrequency $\omega$ for the same dispersion relation. If to run the eigenfrequency along the real axis, the corresponding coherent tune shift will follow a certain line in its complex plane, the conformal map of the real axis in the complex plane of $\omega$ to the complex plane of $\omc$. This line, $\omc(\omega)$, is conventionally called the stability diagram; the beam is stable if and only if its actual coherent tune shift is below the diagram. 

Certain historical investigations~\cite{VaccaroDiagram} convinced the author of this paper that it would be fair to call the stability diagram the Vaccaro diagram (VD), by name of Vittorio Vaccaro, who found how Nyquist's stability plots can be modified to become an effective tool for the collective beam dynamics. With M\"{o}hl-Sch\"{o}nauer dispersion relation~(\ref{DispEq}), VD is determined by the beam distribution function $f$, the lattice frequency shifts and SC tune shifts, being independent of wakes.

In 2001, M.~Blaskiewicz suggested an original method of solving the Vlasov equation with SC and lattice tune spread~\cite{Blaskiewicz:PRAB2001}. The method was free from hidden assumptions of Merle-M\"{o}hl approach, being, arguably, more complicated and less transparent in the computations. The method allowed to make conclusions regarding SC effects for the stability diagram. In case of the chromatic lattice tune spread, the diagram of a Gaussian beam essentially shifted to the left by one half of the maximal SC tune shift. In case of the octupole nonlinearity M.~Blaskiewicz found that focusing octupoles are much more beneficial for LD than defocusing ones, confirming the same conclusion by D.~M\"{o}hl~\cite{Mohl:1995gb}.   

The first analytical attempt to build VD for Eq.~(\ref{DispEq}) was presented in 2004 by E.~Metral and F.~Ruggiero~\cite{Metral:2004qw}. Namely, they suggested a solution of the dispersion relation with SC and octupolar nonlinearity, where, instead of the coasting beam term $n\domi$, the bunched beam term $k \oms$ was put, with $\oms$ as the synchrotron frequency and $k$ as the head-tail mode number. Such extension of the coasting beam theory to the bunched case was, however, left unexplained both in the paper itself and the references it suggested for that matter, including Ref.~\cite{Mohl:1995gb}. As it became more clear later, this extension works reasonably well only when the SC term could be safely omitted. However, the actual merit of the paper was not in its applicability to bunched beams with SC, but in its analytical building of VD for coasting beams with octupoles, nonlinear SC, and insignificant revolution frequency spread, $n\domi =0$. It was confirmed, in particular, that the octupole sign becomes crucial for strong SC; namely, the focusing octupoles are much preferable. The reason is that the octupoles affect mostly the tail particles, so the collective frequency is barely touched by them. Landau damping requires resonant particles, i.e. those which individual tunes are the same as the collective one. Space charge moves the incoherent tunes down, and does almost nothing for the collective tunes, thus killing LD. Thus, to restore the latter, the incoherent tunes have to be moved up to provide higher population of the resonant particles, so the octupoles have to be focusing. It was also shown in this reference that SC can be beneficial if it shifts VD on top of the coherent tune, which would be outside (on the left) of VD without space charge. For very strong SC, it meant that it is detrimental since it shifts the stability diagram far on the left.

Among multiple reasonable features, Vaccaro diagrams of Ref.~\cite{Metral:2004qw} showed a strange one: for defocusing octupoles, there was a kink point of the curve at the real axis, which prevented the line from going to the lower half-plane, $\Im \omc <0$. The kink point looked strange, since VD should be analytical by the definition. 

In 2006, D. Pestrikov published an article~\cite{Pestrikov:2006} where a similar problem was solved, but instead of the kink point, the diagram smoothly continued to the lower half-plain, thus demonstrating Landau antidamping. Later that year Landau antidamping was confirmed by K.Y. Ng for the same model as Metral and Ruggiero proposed~\cite{MetralNg2006}. On the ground of these findings, the kink point of Ref.~\cite{Metral:2004qw} was dismissed as a mistake of the sign. Due to this, however, another problem appeared: at certain conditions, a Gaussian-like beam with time-independent Hamiltonian started looking unstable even when the coherent tune shift $\omc$ suggested a decay of the mode. Next year Pestrikov published another paper~\cite{Pestrikov:2007}, presenting ``a self-consistent model'' which showed no antidamping, contrary to the Merle-M\"{o}hl model; he expressed doubt in the validity of the latter. 

This doubt was enhanced to a stronger claim by V.~Kornilov, O.~Boine-Frankenheim, and I.~Hofmann in their publication of 2008~\cite{Kornilov:2008zz}. First, they confirmed that VD of Eq.~(\ref{DispEq}) indeed yields Landau antidamping for defocusing octupoles. Second, they supported this confirmation by macroparticle simulations within the {\it frozen field} model, equivalent to Merle-M\"{o}hl approach. Third, they ran self-consistent macroparticle simulations for the same conditions, and saw no antidamping. From this, they concluded ``that antidamping can be related to the non-self-consistent treatment of nonlinear space charge in the simulations and also in the dispersion relation.'' 

At that stage, several issues remained unresolved for coasting beams. First, it was not clear if Landau antidamping is ever possible for Gaussian-like beams with SC, octupoles and chromaticity. Second, with evidence of incorrectness of Merle-M\"{o}hl analytical approach at certain cases, it was not clear if their equation could be ever used at all, and under what conditions. Third, no analytical formulas for the instability thresholds were yet obtained. These issues were addressed in Ref.~\cite{BurLebPRAB2009}.

A possibility of Landau antidamping was denied there as contradicting to the Second Law of Thermodynamics. Indeed, a beam with real coherent tune shift $\omc$, corresponding to imaginary transverse impedance, i.e. to zero energy losses, can be described by energy-preserving time-independent Hamiltonian, so the growing coherent oscillations might take energy from the incoherent degrees of freedom only. For a Gaussian beam it would mean a perpetuum mobile of the second kind, forbidden by the Second Law. Landau antidamping, demonstrated for some parameters by Merle-M\"{o}hl dynamic system~(\ref{MMEq}), is caused by the non-Hamiltonian character of its SC term. Specifically, the term $\propto \omsci \xbar $ is non-Hamiltonian unless all the SC frequency shifts are identical within the beam slice. Having said that, it is important to stress that the Merle-M\"{o}hl equation of motion with real coherent tune shift, $\Im \omc=0$, may lead to Landau antidamping only if the incoherent spectrum $\omsci + \domxi$ reaches a local maximum in the action space, which may happen for a defocusing octupole. Although the equation is not Hamiltonian, for monotonic spectra $\omsci + \domxi$ all its van Kampen eigenfrequencies with real coherent tune shift are real as well, no unphysical dissipation is introduced.   

How reliable is Eq.~(\ref{MMEq}) for LD computation for the monotonic spectra?  When the SC tune shifts depend on the transverse actions, as they normally are, the defect of the model still should not play a role, if the slices were sufficiently rigid in their transverse oscillations. In this case only the tail particles would be responsible for LD, so the energy transfer to them could be reasonably approximated with the rigid core model. To see when the core is really rigid, note that if the lattice tune shifts are small with respect to the tune separation,  
\begin{equation}
|\domxi + n\domi| \ll |\omc - \omsci|\,,
\label{SSCcond1}
\end{equation}
the particles move together with the related centroids, $x_i \approx \xbar$, as it follows from Eq.~(\ref{eigen2}). Thus, if the SC is so strong that this condition is satisfied for the majority of particles, the slices oscillate almost without distortions, since almost all the particles oscillate almost identically to their centroids; so the rigid-slice approximation is justified. Luckily, for many low- and medium-energy machines, typical SC tune shifts are much larger than the imaginary part of the coherent tune shifts, $|\omsci | \gg \Im \omc$, so stabilization is achieved at such a small lattice nonlinearity that Eq.~(\ref{SSCcond1}) is satisfied, justifying Merle-M\"{o}hl equation. In this case of strong SC, instability thresholds were explicitly found in Ref.~\cite{BurLebPRAB2009} for Gaussian beam, both for octupolar and chromatic frequency spreads. Recently, this method was extended to electron lenses; Landau damping rate introduced by a Gaussian e-lens for a coasting beam with SC was analytically estimated and presented in Ref.~\cite{LebECooler}.

\section{\label{SecBunch} Bunched Beams}

The coherent spectrum of bunched beam with SC was presented for the first time by M.~Blaskiewicz in 1998~\cite{blaskiewicz1998fast} within a simple model of an air-bag bunch in a square potential well. For a delta-wake, the eigenfrequencies were found to be same, as Eq.~(\ref{2stream}) for the two-stream coasting beam, with the substitution $n \domo \rightarrow k \oms$, where $k=0, 1, 2,...$ is the mode counter and $\oms$ is the synchrotron frequency. 
A new and rather surprising mathematical result of M.~Blaskiewicz~\cite{blaskiewicz1998fast} showed suppression of the transverse mode coupling instability (TMCI) by SC; the wake threshold was demonstrated to grow with SC tune shift, linearly at the strong SC limit, $|\omsc | \gg \oms$. This result was obtained for exponential wakes and the ABS model (Air-Bag, Square-well), so a question was raised about the sensitivity of this unexpected result to the details of the wake, potential well and bunch distribution. Also, it was not clear if there was any limit to this growth of the instability threshold. An explanation of this growth at moderate SC was suggested to the author by V.~Danilov~\cite{SlavaComm1998} and reproduced in Ref.~\cite{Ng:1999fy}. Without SC, TMCI typically results from crossing of the head-tail mode $0$, shifted down by the wake, and the mode $-1$, not shifted as much. Space charge, on the contrary, does not influence the mode $0$ and shifts down the mode $-1$, thus moving their coupling point to higher intensity.      

In the year of 2009, when it was understood that Merle-M\"{o}hl approach of rigid slices is justified for sufficiently strong space charge, it was applied to bunched beams by the author~\cite{burov2009head}. Under the condition of SC tune shift being much stronger than all other tune shifts and spreads, as well as the synchrotron tune (strong space charge, SSC), an ordinary linear integro-differential equation was derived for the bunch modes for an arbitrary potential well, driving and detuning wakes, longitudinal and transverse bunch distribution functions. Later that same year V.~Balbekov published a paper~\cite{Balbekov:2009PRAB} with an alternative derivation of the SSC equation, which result differed from mine. After checking his derivation and rechecking mine, I found an algebraic error in my calculations, and derived my ultimate form of the SSC mode equation, which agreed with Balbekov's result, suggesting a slightly more compact form in the erratum~\cite{Burov2009PRABErratum}, 
\begin{equation}
i\frac{\pd \xbar}{\pd t} + \frac{1}{\omsc} \frac{\pd }{\pd s} \left( u^2 \frac{\pd \xbar}{\pd s} \right)= \mathbb{W} \xbar + \mathbb{D} \xbar \,.
\label{MySSCEq}
\end{equation}
Here $\omsc = \omsc(s)$ is the SC frequency shift averaged over the transverse actions at every position $s$, $u^2=u^2(s)$ is the local rms spread of the longitudinal velocities, $u^2 \equiv \left < R^2 \domi^2 \right>$, while $\mathbb{W}$ and $\mathbb{D}$ are conventional driving and detuning wake linear integral operators~\cite{BurovDanilovPRL1999}; in more details see~\cite{Burov2009PRABErratum}. The equation is complemented by zero-derivative boundary conditions, $\pd \xbar /\pd s =0$ at the bunch edges or at $s=\pm \infty$. For the eigenfunctions, the time derivative has to be substituted by the sought-for eigenfrequency $\nu$, i.e. $i \pd /\pd t \rightarrow \nu$. Without wakes, this equation leads to the Blaskiewicz-type collective spectrum, $\nu_k \simeq k^2 \oms^2/\omsc$. The mathematical elegance of Eq.~(\ref{MySSCEq}) has its price: missing is the Landau damping, which required additional ideas and computations.  

Analytical estimations for LD at SSC were also suggested in Ref.~\cite{burov2009head, Burov2009PRABErratum} for weak head-tail cases, when the wake does not influence the eigenfunction much. Contrary to coasting beams, it was found that there is an intrinsic LD, caused by the longitudinal variation of the SC tune shift only, even without any lattice tune spreads. The physical mechanism of the dissipation was associated with a break of the slice rigidity at the bunch edges, where the SC is not strong any more. The slice softening at the bunch edges opens a way for the energy transfer to the incoherent degrees of freedom. According to the related estimation, the intrinsic LD rate $\Lambda_k$ at SSC was found to be a steep function of the SC parameter $q \equiv \omsc/\oms$ and the positive mode number $k$, $\Lambda_k \simeq k^4 \oms/q^3$; the SSC assumes $q \gg 2k$.  Six years later these analytical results for SSC eigenfunctions and LD rates were fully confirmed in Synergia macroparticle simulations by A.~Macridin et al.~\cite{MacridinPRAB2015}, where the intrinsic LD rates were shown to have their maxima at $q \simeq 2k$. A more subtle case of {\it parametric Landau damping} was treated by A.~Macridin et al. in Ref.~\cite{MacridinPRAB2018} by means of analytical modeling and macroparticle simulations. Analytical estimations of octupoles-related LD suggested in Ref.~\cite{burov2009head, Burov2009PRABErratum} are still waiting for at least numerical verifications; nothing yet has been published in that matter. Octupoles, however, are rather inefficient for LD, which requires significant nonlinearity inside the beam, not far outside, as octupoles provide. That is why a better instrument for LD is an electron lens, at least as thin as the beam. Such e-lenses are able to provide LD without deterioration of the dynamic aperture, as it was pointed out by V.~Shiltsev et al.~\cite{ShilAlexBurValPRL2017}. Estimations of e-lens-caused LD rates for bunches with SC were suggested by Yu.~Alexahin, A.~Burov and V.~Shiltsev in 2017~\cite{AlexahinBurShil2017rep}.    

With the wake taken into account, the Blaskiewicz' result of linear growth of the TMCI wake threshold was confirmed in a series of publications, see Refs.~\cite{ZolkinBurovPandeyPRAB2018, BalbekovPRAB2019} and references therein. A hidden obstacle with this problem, sometimes caused  misleading results, was realized by V.~Balbekov~\cite{balbekov2017transverse}, who showed that convergence of the expansion of the sought-for eigenfunction over the zero-wake basis degrades with SC, requiring more and more terms for higher SC parameter $q$. The physical reason of the convergence worsening was recently found by the author of this paper; it is associated with the head-to-tail amplification, or the {\it convective instabilities} driven by wakes at SSC. When eigenfunctions are significantly amplified, their expansion over any even basis cannot be of a good convergence. Clearly manifest subsiding of the instability with SC was presented in the two-particle model of Ref.~\cite{ChinTwoPartPRAB2016}. 

With the theoretical proof of TMCI vanishing with SC, two problems became rather obvious, one experimental and the other theoretical. The former consisted in a reasonable agreement of the transverse instability at CERN SPS with no-SC theory, while SC tune shift was very strong there, especially with the old lattice~\cite{Bartosik:2013qji, MetralFNAL2018}. The latter problem was related to the linear growth of the wake threshold with SC. Due to this feature, the bunch should be stable up to infinite intensity, as soon as its emittance is low enough, which did not sound as a reasonable statement. The resolution of both problems was presented by the author two years ago~\cite{BurovConvectivePRAB2019}. The main idea, already mentioned above, was that, while moving out TMCI, SC moves another instability in its place, a convective one. Contrary to TMCI, which is an absolute instability, i.e. has nonzero growth rate, the convective instabilities grow not in time, but in space, from head to tail~\cite{LifPitKin}. This head-to-tail amplification increases exponentially with bunch intensity, resulting in one or another physical limit, set by lattice nonlinearity, beam loss or feedback. When the amplification is large, even a tiny feedback from tail to head may be sufficient to close the loop and turn the convective instability into an {\it absolute-convective} one, like those with a microphone close to its loudspeaker. Such a feedback may be presented with a bunch-by-bunch damper, coupled-bunch wakes, or a halo of the same bunch. Here a question may be asked, why is the halo needed for the feedback? Why can core particles not play this role, when they move to the bunch head with their high transverse amplitudes acquired at the tail? The answer is that due to strong SC, the bunch slices are rigid, as it was discussed in the previous section. Strong SC means that all tune shifts are small compared with the SC tune shift, so intra-slice degrees of freedom cannot be excited, and thus the tail particles do not preserve their large amplitudes while moving to the bunch head; instead, they just follow the existing spacial pattern of the rigid-slice oscillations. 

However, what is impossible for the bunch's core, might work for its halo, which SC tune shift is smaller, so the halo slices can be soft, providing a tail-to-head feedback. At strong SC, this feedback would be small due to the low population of the halo, but, if the convective amplification is large enough, even a small feedback could be sufficient to ignite the absolute-convective instability, as it was suggested and modeled in Ref.~\cite{Burov:2018rmx}. Apparently, the same effect is responsible for the non-monotonic behavior of the wake threshold on the SC parameter reported by Yu. Alexahin at this workshop~\cite{AlexahinZermatt2}. A good agreement of his analytically calculated highly convective mode with the pattern of oscillations seen by A.~Oeftiger in macroparticle simulations for the same conditions also deserves to be mentioned. 

It is already clear, that convective instabilities constitute a common obstacle for high intensity circular machines of low and medium energy, where SC is significant; they definitely take place at CERN Booster, PS and SPS rings, as well as at the Fermilab Booster. That is why it is important to understand how they behave together with other factors of beam dynamics. Transverse instabilities of a bunch with SC, wake and damper were considered in Ref.~\cite{Burov:2018zwb}. 
In Ref.~\cite{KornilovPS2013}, measurements of a microwave instability at transition crossing in PS were reported; the instability was characterized as {\it convective}. Recently an analytical model for it was proposed~\cite{Burov:2019myg} by means of Eq.~(\ref{MySSCEq}). A simple threshold formula derived there was found to be in good agreement with the data of Refs.~\cite{KornilovPS2013, MiglioratiPRAB2018}. A statement made in Ref.~\cite{KornilovPS2013} that ``The bunch parameter measurements demonstrate... that the space-charge effect does not affect the instability thresholds'' does not actually contradict to rather weak dependence of the threshold bunch intensity $N_\mathrm{th}$ on the transverse emittance, $N_\mathrm{th} \propto \epsilon_\perp^{1/4}$ of Ref.~\cite{Burov:2019myg}, since the limited range of the emittances examined in Ref.~\cite{KornilovPS2013} and the measurement errors do not allow to resolve rather weak dependence on the emittance on the ground of this set of measurements alone~\cite{KornilovPrivate2020}.

%

%

\begin{acknowledgments}

I am indebted to Elias Metral for detailed discussions in May 2018 which clearly showed me the contradiction between the SPS observations and theory of the vanishing TMCI, as the latter was understood at that time. I am also grateful to Elias for his support in researching the historical issues that I tried to reflect here. I appreciate the clarifications of Vladimir Kornilov concerning his PS measurements and remarks of Valeri Lebedev which helped me prepare this paper.

Fermilab is operated by Fermi Research Alliance, LLC under Contract No. DE-AC02-07CH11359 with the United States Department of Energy.

\end{acknowledgments}

\bibliography{bibfile}			
\end{document}